\documentclass[reprint,fleqn,superscriptaddress,twocolumn,showpacs,
amsmath,amssymb,prb,aps,longbibliography]{revtex4-1}
\usepackage[all]{xy}
\usepackage{gensymb}
\usepackage{graphicx,psfrag,times,epsfig,color}
\usepackage{verbatim,natbib}

\usepackage{color}
\usepackage{makeidx}
\usepackage{amsmath}
\usepackage{bm}
\usepackage{amsfonts}
\usepackage{amssymb}
\usepackage{hyperref}

\begin{document}

\title{Superconducting correlations induced by charge ordering in
cuprate superconductors and Fermi arc formation}
\author{E.V.L. de Mello}
\email[Corresponding author: ]{evandro@if.uff.br}
\affiliation{Instituto de F\'{\i}sica, Universidade Federal Fluminense, 24210-346 Niter\'oi, RJ, Brazil}

\author{J.E. Sonier}
\affiliation{Department of Physics, Simon Fraser University, Burnaby, British Columbia V5A 1S6, Canada.}
\affiliation{Canadian Institute for Advanced Research, Toronto, Ontario M5G 1Z8, Canada.}

\begin{abstract}
We have developed a generalized electronic phase separation model of high-temperature
cuprate superconductors that links the two distinct energy scales of the superconducting
(SC) and pseudogap (PG) phases via a charge-density-wave (CDW) state. We show that
simulated electronic-density modulations resembling the charge order (CO) modulations
detected in cuprates intertwine the SC and charge orders by localizing charge and
providing the energy scale for a spatially periodic SC attractive potential. Bulk
superconductivity is achieved with the inclusion of Josephson coupling between nanoscale
domains of intertwined fluctuating CDW and SC orders, and local SC phase fluctuations
give rise to the Fermi arcs along the nodal directions of the SC gap. We 
demonstrate the validity of the model by reproducing the hole-doping dependence of 
the PG onset temperature $T^*$, and the SC transition temperature $T_c$ of 
${\rm YBa_2Cu_3O_y}$ and ${\rm Bi_{2-y}Pb_ySr_{2-z}La_zCuO_{6+\delta}}$. The results show that the periodicity of the CDW order is controlled by the PG
energy scale, and the hole-doping dependence of the SC energy gap is controlled by the charge ordering free energy.
\end{abstract}

\maketitle

\section{Introduction}\label{intro} 

Experiments using different methods have established the occurrence of short-range,
incommensurate static CDW correlations in a variety of high-temperature SC cuprates
\cite{Chang2012,Ghiringhelli2012,Comin2014,Blanco-Canosa2014,Huecker2014,
DaSilvaNeto2015,Comin2015a,Comin2015b,DaSilvaNeto2014,Fujita2014,Wise2008,Wise2009,
Hoffman2002,Howald2003,Vershinin2004,Hanaguri2004,McElroy2005}. With
the exception of La-based cuprates in which CDW order is accompanied by spin order, the
charge order (CO) observed in different cuprate families appears to be similar. In zero magnetic
field, the CDW order is essentially two-dimensional. The wave vector of the CDW order is
parallel to the Cu-O bond directions along the a and b axes, and decreases in magnitude with
increased charge doping. While much of the experimental data cannot distinguish between
checkerboard (bidirectional) or alternating stripe (unidirectional) CO, recent resonant X-ray
scattering (RXS) experiments on underdoped ${\rm YBa_2Cu_3O_7}$ (Y123)\cite{Comin2015a} 
and an analysis of scanning
tunneling microscopy (STM) data for ${\rm Bi_2Sr_2CaCu_2O_{8+\delta}}$\cite{Hamidian2016} 
indicate that the inter-unit-cell character
is one of segregated or overlapping unidirectional charge-ordered stripes. Furthermore, it has
been found that the CDW order possesses a $d$-wave intra-unit-cell symmetry with the modulated
charge primarily on the O-$2p$ orbitals linking the Cu atoms
\cite{Comin2015b,DaSilvaNeto2014,Fujita2014}. Since the SC order 
parameter also
has $d$-wave symmetry, this local charge or bond order symmetry supports theoretical proposals
that suggest the charge and SC order parameters are intimately intertwined. Some attribute the $d$-
wave CO symmetry to quasiparticle scattering by antiferromagnetic (AF) fluctuations near a
metallic quantum critical point, which also gives rise to the d-wave superconductivity\cite{Sachdev2013,Efetov2013,Hayward2014}.
Alternatively, it has been proposed that CDW order in cuprates is a consequence of a pair-density
wave (PDW) phase, in which the SC order parameter is periodically modulated in space due to
the Cooper pairs having finite momentum\cite{Pepin2014,Fradkin2015,Wang2015a,Wang2015b}.

The aim of our work is to establish a quantitative link between the inter-unit-cell
dependence of the CO resolved by RXS and imaged in real space by STM, and the energy gaps
of the PG and SC phases. Our model is based on an intrinsic propensity for mesoscale electronic
phase separation below an onset temperature $T_{\rm PS}$ that follows the hole-doping dependence of the
PG temperature $T^*$. This picture is similar to that previously advocated by Fradkin and
Kivelson\cite{Fradkin2012}.
We presume the onset of fluctuating CDW order domains at $T^*$ , where STM
measurements on ${\rm Bi_2Sr_2CaCu_2O_{8+\delta}}$ have detected the emergence of charge stripes that extend into
the overdoped regime\cite{Parker2010}. The short-range static CO that has been observed by X-rays at a lower
temperature $T_{\rm CO} \leq T^*$ is assumed to be confined to local regions where fluctuating CDW order
has become pinned by disorder.  Contrary to this assumption, we note 
that in ${\rm HgBa_2CuO_{4+\delta}}$ (Hg1201) CDW order observed by X-ray 
scattering vanishes already well below optimal doping\cite{Tabis2017}. 
This seems to be due to the presence of pairs of interstitial oxygens within the 
same unit cell specific to Hg1201. Although not captured by our model, it is 
also important to recognize that the pseudogap region marks the
onset of an intra-unit-cell magnetic order\cite{Fauque2006,Li2008},
a true phase transition that modifies ultrasonic waves\cite{Shekhter2013},
an increase in antiferromagnetic correlations\cite{Chan2016}
and global inversion-symmetry breaking\cite{Zhao2017}.

Another important ingredient of our model is the experimental observation that the CDW periodicity is independent of temperature, leading us to surmise that
the CDW periodicity is set by the onset of the PG at $T^*$ . This infers that 
the CDW order is a
consequence of the PG formation. At low doping ($p \leq 0.12$) where $T_{\rm CO}$ decreases with decreasing
doping, CDW order is potentially suppressed by a slowing down of spin 
fluctuations and a
tendency toward static SDW order. Compatible with experimental signatures 
of pairing or SC correlations persisting above
$T_c$\cite{Corson1999,Xu2000,Gomes2007,Yang2008,Dubroka2011,Li2005,Wang2005}, our model shows that CDW order in the PG regime may
induce SC domains that grow and connect to establish bulk superconductivity at $T_c$.

\section{Simulation of the charge-ordered state}

\begin{figure*}
\includegraphics[height=8cm]{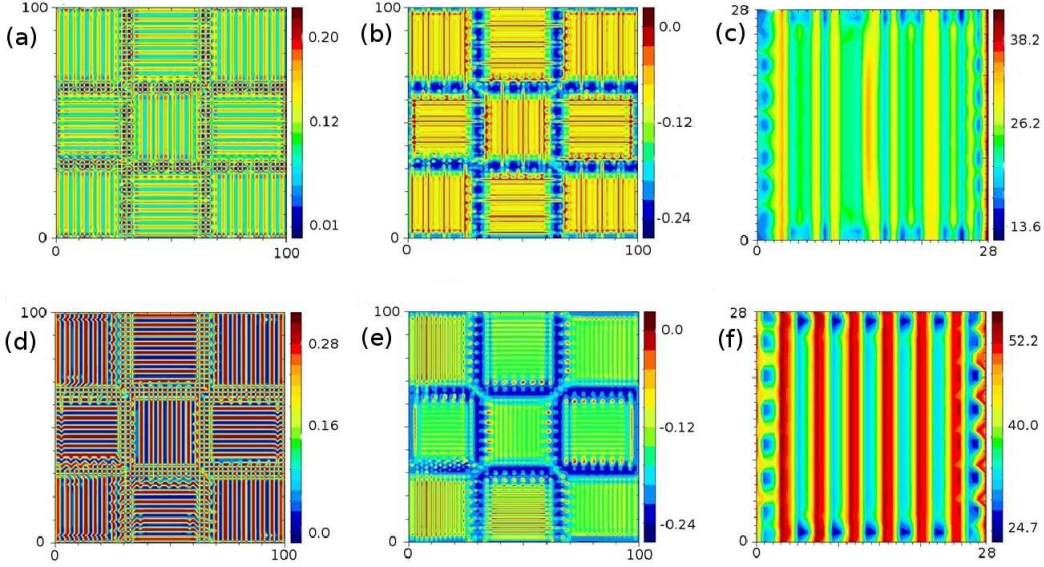}
\caption{Contour plots of the electronic density $p$({\bf r})
calculated on a square lattice of $100 \times 100$ unit cells, with 
average charge densities of $p = 0.12$ in (a) and $p = 0.16$ in (d). 
The charge order wavelengths are $\lambda_{\rm CO} = 3.15 a_0$ (a) and 
$\lambda_{\rm CO} = 3.49 a_0$
in (d), corresponding to the charge order wave vectors 
determined by momentum-resolved X-ray
probes\cite{Blanco-Canosa2014,Huecker2014,Comin2015a}. (b), (e) Corresponding spatial dependence of the free-energy potential 
$V_{\rm GL}$({\bf r}). The
periodicity of the potential manifests in the periodic modulations of the charge density. (c), (f)
Results of calculations of the d-wave pairing potential $\Delta_d$({\bf r}) 
displayed for a single domain over a
$28 \times 28$ unit cell area (in meV unit). The spatial average value of the pair potential
$\langle \Delta_{d}({\bf r})\rangle$ is 25.5 meV at $p =
0.12$ in (c), and 43.8 meV at $p = 0.16$ in (f).}
\label{fig1}
\end{figure*}

Our approach is to first simulate spatial modulations of the electronic structure resembling experimentally resolved inter-unit-cell CO modulations, using the time-dependent Cahn-Hilliard (CH) differential equation\cite{Cahn1958}. Besides generating the desired CDW order, the CH approach yields the associated free-energy modulations, which we assume scales with a periodic attractive potential in the subsequent SC calculations. The starting point is the introduction of a time-dependent conserved order parameter associated with the local electronic density, 
$u({\bf r}, t) = [p({\bf r}, t) - p]/p$, where $p$ is the average charge density and 
$p({\bf r}, t)$ is the charge density at a position ${\bf r}$ in the plane. The Ginzburg-Landau (GL) free energy density of the system is of the form

\begin{figure*}
\includegraphics[height=8cm]{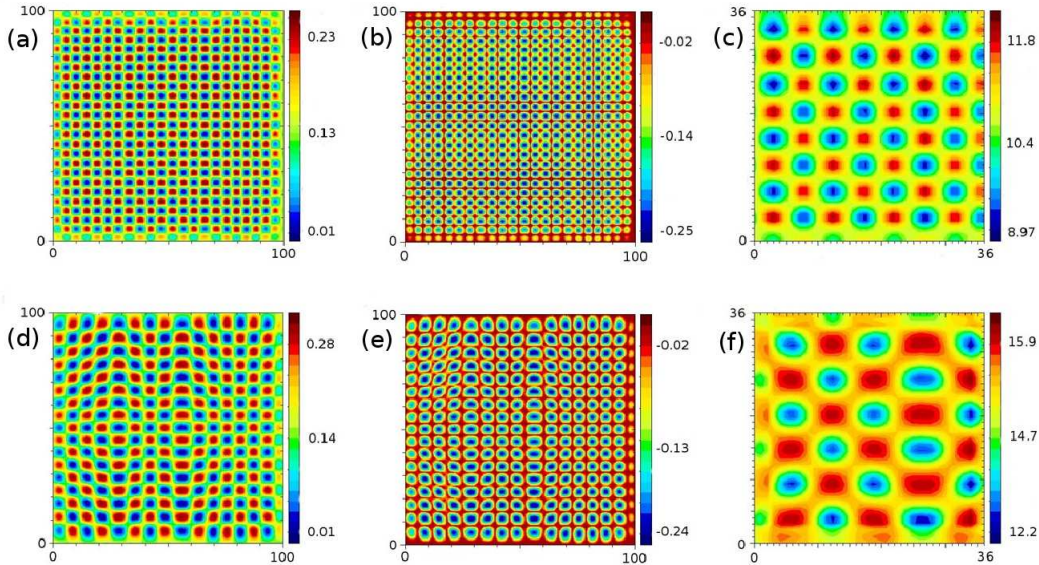}
\caption{Contour plots of the electronic
density $p$({\bf r}) calculated on a square lattice of 
$100 \times 100$ unit cells, assuming average charge
densities of $p = 0.126$ in (a) and $p = 0.16$ in (d). 
The charge order wavelengths are $\lambda_{\rm CO} = 4.5a_0$ in a
and $\lambda_{\rm CO} = 6.2a_0$ in (d), matching the checkerboard wavelength of 
the STM conductance maps of
underdoped ($T_c = 25$ K; $p = 0.126$) and optimally-doped (Pb, La)-Bi2201 
in Ref. \onlinecite{Wise2008}. (b), (e),
Corresponding spatial dependence of the free-energy potential 
$V_{\rm GL}$({\bf r}). (c), (f), Results of calculations
of the $d$-wave pairing potential $\Delta_d$({\bf r}) displayed over 
a $36 \times 36$ unit cell area (in meV unit). The spatial average
value of the pair potential $\langle \Delta_{d}({\bf r})\rangle$ is 9.2 meV at 
$p = 0.126$ in (c), and 15.8 meV at $p = 0.16$ in (f).}
\label{fig2}
\end{figure*}

\begin{equation}
f(u)= {{\frac{1}{2}\varepsilon |\nabla u|^2 +V_{GL}(u,T)}},
\label{FE}
\end{equation}
where  ${V_{\rm GL}}(u,T)= -\alpha [T_{\rm PS}-T] u^2/2+B^2u^4/4+...$ is a
double-well potential that characterizes the electronic phase separation 
below $T_{\rm PS}$. The parameters
$\alpha$ and $B$ are constants, and 
$\varepsilon$ controls the spatial separation of the charge-segregated patches.
The CH equation obtained from
the continuity equation for the local free energy current density 
${\bf {\rm J}} = -{\rm M}\nabla \mu$ (where M is the charge
mobility and $ \mu = \partial f/\partial\mu$ is the chemical potential). 
Its solution is described in the Apendix I
on charge ordering simulations. For each time step the CH equation is solved for
$u({\bf r}, t)$,
and $p({\bf r}, t)$ is obtained. We adjust the parameters of the free energy such that when the periodicity
of $p({\bf r}, t)$ matches that of the experimentally observed CDW order, the calculation is stopped and
the solution is taken to be the spatially-dependent static electronic density 
$p({\bf r})$. Since the method
described here does not generate an intra-unit-cell CO symmetry, it is applicable to systems that
have SC pairing and CO symmetries other than d-wave.

Figure \ref{fig1}(a) shows a simulation of alternating planar domains of $90^0$-rotated charge stripes with
an intra-domain periodicity compatible with RXS data for detwinned Y123 at 
$p = 0.12$[\onlinecite{Comin2015b}]. Within
each domain the CO wavelength is $\lambda_{\rm CO} = 3.15a_0$, where the in-plane lattice constant is $a_0 = 3.85 \AA$. The CO wavelength in Y123 measured by various X-ray methods increases with increased hole
doping\cite{Blanco-Canosa2014,Huecker2014,Comin2016} . We have used such data to generate similar CO striped patterns for Y123 at $p = 0.16$
(Fig. \ref{fig1}(d)) and $p = 0.09$.

STM differential conductance maps for optimally-doped ($T_c = 35$ K) and 
underdoped ($T_c
= 32$ K and $T_c = 25$ K) ${\rm Bi_{2-y}Pb_ySr_{2-z}La_zCuO_{6+\delta}}$ 
[(Pb, La)-Bi2201] samples\cite{Wise2008,Wise2009} exhibit checkerboard
patterns (indicative of the simultaneous presence of both CDW domains) 
with $6.2 ± 0.2a_0$ , $5.1 ± 0.2a_0$ , and $4.5 ± 0.2a_0$ unit cell 
($a_0 = 3.83 \AA$) periodicity, respectively. The increase in the CO
wavelength with increased hole doping agrees well with X-ray scattering and STM measurements
on Bi2201 without cation substitutions\cite{Blanco-Canosa2014} . We have used the following formula from Ref. \onlinecite{Schneider2005} to
calculate the average number of holes per Cu for the 
(Pb, La)-Bi2201 samples in Ref. \onlinecite{Wise2008}:
$T_c /T^{\rm max}_c = 1 - 250(p - 0.16)^2$ , where $T^{\rm max}_c = 35$ K. This calculation yields $p = 0.126$ and $p =
0.141$ for the two underdoped samples. Figs. \ref{fig2}(a) and \ref{fig2}(d) display CO checkerboard patterns
simulated by the CH equation that resemble the STM differential conductance 
maps for (Pb, La)-Bi2201 at $p = 0.126$ and $p = 0.16$.

The CO periodicity is manifest in the spatial dependence of the free-energy potential
$V_{\rm GL}({\bf r})$, shown in Figs. \ref{fig1}(b) and \ref{fig1}(e), 
and Figs. \ref{fig2}(b) and \ref{fig2}(e). The central assumption 
in our model is that
by confining charge, a fluctuating CDW periodic potential that scales with
$V_{\rm GL}({\bf r})$ mediates the
attractive two-body SC interaction. In particular, we assume the 
fluctuating periodic potential has
the same periodicity as the static CO detected experimentally and 
has a time-averaged potential
well depth that is proportional to the depth of the static periodic 
potential. In what follows, we
make the approximation $~ \langle V_{\rm GL}({\bf r})\rangle$, where $\langle 
V_{\rm GL}({\bf r})\rangle$ is the spatial average of $V_{\rm GL}({\bf r})$ 
over a $100 \times 100$ unit cell area.

\section{Superconducting calculations in the charge-ordered state}

Next we use the free-energy simulations and experimentally determined input 
parameters for the optimally-doped compounds to deduce the SC energy gap 
$\Delta_{\rm SC}$, the pseudogap $\Delta_{\rm PG}$, $T_c$ and $T^*$ 
for the underdoped samples. To derive the local SC gap we solved the Bogoliubov-deGennes (BdG) equations via self-consistent calculations based on a Hubbard Hamiltonian [Eq. \ref{Ham}]. The calculations were performed for a sub-lattice about the center of the simulated charge density maps, using periodic boundary conditions and governed by self-consistent conditions for a spatially-varying d-wave pairing potential $\Delta_{d}({\bf r})$ and hole density $p({\bf r})$ 
[Eq. \ref{Deltad}) and \ref{dop}]. We find that the spatial-average 
$\langle \Delta_{d}({\bf r})\rangle$ decreases with a reduction of $p$ 
(below $p = 0.16$), but increases with decreasing $\lambda_{\rm CO}$. 
The latter behavior is because as the two holes are forced closer 
together by the narrower confining potential the binding energy of the 
two-body interaction increases. The results on the CuO$_2$ plane shown in 
Figs. 1(c), 1(f), 2(c) and 2(f) indicate that in our approach 
the PDW is a consequence of the CDW.

\begin{figure}
\centering
\begin{minipage}{0.46\textwidth}
\includegraphics[width=7.0cm]{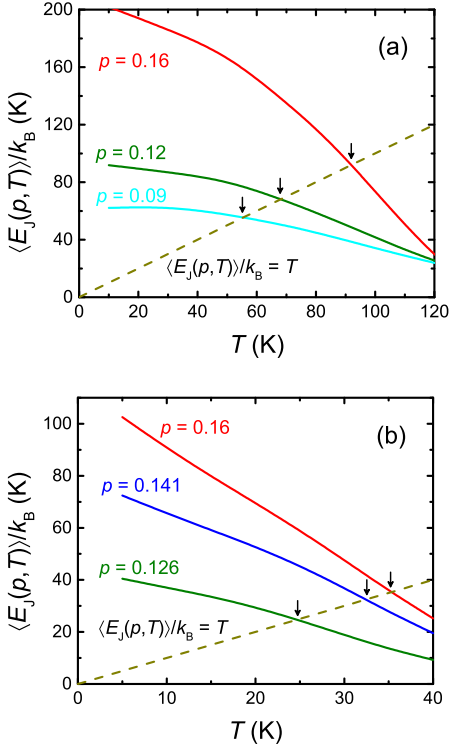}
\end{minipage}
\caption{(Color online)
Calculated values of the superconducting transition temperature. (a,b) 
The temperature dependence of the average Josephson coupling energy 
$\langle E_{\rm J}(p, T)\rangle$ (divided by
Boltzmann constant $k_{\rm B}$) for Y123 at $p = 0.09, 0.12$, and 0.16 in (a), 
and (Pb, La)-Bi2201
at $p = 0.126, 0.141$, and 0.16 in (b). The values of $T_c$ correspond to the intersections of the
dashed straight line $\langle E_{\rm J}(p, T)\rangle/k_{\rm B} = T$
with the $\langle E_{\rm J}(p, T)\rangle/k_{\rm B}$ curves and are marked by
the arrows.
}
\label{fig3}
\end{figure}

\begin{figure}
\includegraphics[height=7.0cm]{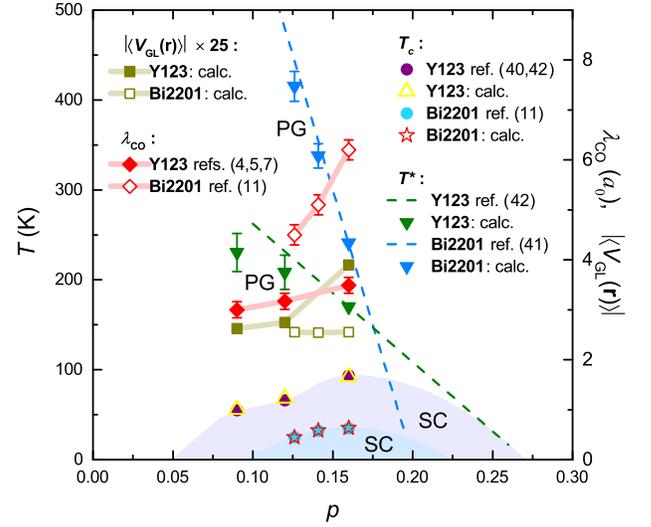}
\caption{Comparison of experimental and calculated values of 
$T_c$ and $T^*$ versus hole doping. Also
shown is the experimentally determined doping dependence of the charge order wavelength  $\lambda_{\rm CO}$
for both compounds, as well as the doping dependence of the calculated 
absolute value of the
spatial average of the free-energy potential $|\langle V_{\rm GL}({\bf r})\rangle|$.
For display purposes 
$|\langle V_{\rm GL}({\bf r})\rangle|$ is shown multiplied by a factor of 25.
}
\label{fig4}
\end{figure}

The values of  $~ \langle V_{\rm GL}({\bf r})\rangle$ at optimal doping were
multiplied by a scaling factor, such that the calculations (Figs. \ref{fig1}(f) 
and \ref{fig2}(f)) generate an average value of the pairing potential $\langle \Delta_{d}({\bf r})\rangle$ 
for $p = 0.16$ that is close to the experimentally estimated value of the 
low-temperature SC gap $\Delta_{\rm SC}$. To calculate 
$\langle \Delta_{d}({\bf r})\rangle$ for the underdoped samples 
(Tables \ref{table1} and \ref{table2}), this same scaling factor was subsequently applied 
to the respective values of $~ \langle V_{\rm GL}({\bf r})\rangle$. 
For (Pb, La)-Bi2201, 
the value of $~ \langle V_{\rm GL}({\bf r})\rangle$ varies little with doping, and hence $p$ and $\lambda_{\rm CO}$ are responsible for the hole-doping dependence of $\langle \Delta_{d}({\bf r})\rangle$. Experimental estimates of the SC gap for the $p = 0.126$
and 0.141 samples are not reported, but the calculated values of 
$\langle \Delta_{d}({\bf r})\rangle$ for the underdoped samples (Table \ref{table2}) 
roughly follow the trend expected if the ratio $\Delta_{\rm SC}/k_{\rm B}T_c$
is independent of $p$. In contrast to (Pb, La)-Bi2201, the doping dependence 
of $\lambda_{\rm CO}$ in Y123 is weaker, and the CH simulations of charge stripes are characterized by a significant change in $\langle V_{\rm GL}({\bf r})\rangle$ 
with doping (Fig.\ref{fig4}). Consequently, the depth of the periodic potential plays an important role in the calculation of the doping dependence of 
$\langle \Delta_{d}({\bf r})\rangle$ for Y123. The calculated values of $\langle \Delta_{d}({\bf r})\rangle$ at p = 0.09 and 0.16 agree well with an empirical 
relation for $\Delta_{\rm SC}(p)$ that describes a number of high-$T_c$ 
cuprate superconductors\cite{Huefner2007}. The calculated result 
at $p = 0.12$ falls below 
this universal curve, which is consistent with the well-known plateau of 
$T_c(p)$ for Y123 near 1/8 hole doping.

An estimate of $T_c$ is obtained by self-consistently solving the BdG equations 
with a temperature-dependent GL potential
\begin{equation}
 V(T,p) = V(0,p)[1-T/T_{\rm PS}(p)]^2,
\label{V(T)}
\end{equation}
where we take $V(0,p) \approx \langle V_{\rm GL}({\bf r},p)\rangle$ and 
$T_{\rm PS}(p)$, the onset of phase separation transition, is taken to be
equan to $T^*(p)$ in the calculations. Because of the BdG approach and the above equation, the value of $\langle \Delta_{d}({\bf r},T)\rangle$  
decreases with increasing temperature, but the it remains finite in many regions of the system for a significant range of temperature above Tc.  This is consistent with the body of experimental results on cuprates mentioned earlier that are suggestive of persisting SC correlations above $T_c$\cite{Corson1999,Xu2000,Gomes2007,
Yang2008,Dubroka2011,Li2005,Wang2005}. Typical 
$\langle \Delta_{d}({\bf r},T)\rangle$ plots of three Y123 
compounds are shown in Fig. A2. Next we assume that bulk superconductivity 
is achieved via Josephson coupling between different closely spaced patches of intertwined  
CO and SC pairing. We assume there is SC phase coherence within the patches, and that there are many such closely spaced SC domains slightly above 
$T_c(p)$ forming junctions with an average tunnel y proportional to the normal-state resistance immediately above $T_c(p)$. As explained previously\cite{DeMello2014}, 
for a $d$-wave superconductor in single-crystal form it is sufficient to use the following relation for the average Josephson coupling energy

\begin{equation}
 \langle E_{\rm J}(p,T)\rangle = {\frac{\pi \hbar \langle 
 \Delta_{d}({\bf r},T)\rangle}{2 e^2 R_{\rm n}(p)}} \tanh\left[
 {\frac{\langle \Delta_{d}({\bf r},T)\rangle}{2 k_{\rm B}T} } \right]
\label{EJ(T)}
\end{equation}

where $\langle \Delta_{d}({\bf r},T)\rangle = \sum_{i}^{N} \langle \Delta_{d}
(r_i,p,T)\rangle / N$ is the pairing potential. The quantity $R_{\rm n}(p)$ 
is proportional to the normal state in-plane resistivity $\rho_{ab}(p, T \geq T_c)$.
In what follows we assume $R_{\rm n}(p)$ for the optimally-doped compounds, and in the 
case of (Pb, La)-Bi2201 use experimental values of the in-plane resistivity 
ratio $\rho_{ab}(p)/\rho_{ab}(p = 0.16)$ to calculate $R_{\rm n}(p)$ for the 
underdoped samples. Since the relationship between $T_c/T_c^{\rm max}$ and 
the hole concentration $p$ for (Pb, La)-Bi2201 is the same as for La-doped 
Bi2201\cite{Schneider2005}, we have used available resistivity data for La-doped Bi2201\cite{Ono2003} in our calculations shown in Table \ref{table2}. For orthorhombic Y123, we instead used experimental values of the $b$-axis resistivity ratio $\rho_b(p)/\rho_b(p = 0.16)$ from Ref. \onlinecite{Segawa2001} 
to estimate $R_{\rm n}(p)/R_{\rm n}(p = 0.16)$.

As the temperature is lowered below $T^*$, thermal fluctuations diminish and 
long-range phase coherence between the individual SC domains is established when 
$k_{\rm B}T \approx \langle E_{\rm J}(p, T)\rangle$. The temperature $T$ at 
which this occurs defines the bulk critical temperature $T_c$. 
Figs. \ref{fig3}(a) and \ref{fig3}(b) show the temperature dependence of $\langle E_{\rm J}(p, T)\rangle$
for both compounds at the different dopings. The intersection of the 
$\langle E_{\rm J}(p, T)\rangle$ curves with the $k_{\rm B}T$ line yields values 
of $T_c$  in good agreement with the actual values for Y123 
and (Pb, La)-Bi2201 (Fig. \ref{fig4} and Tables \ref{table1} and \ref{table2}).

\section{The pseudogap}

Next, we use the free-energy simulations to make a simple estimate of the PG, 
under the assumption that the PG appears due to the mesoscale phase 
separation\cite{Fradkin2012} that creates small domains of CO wavelength below 
$T_{\rm PS} \approx T^*$. For Bi2201 we consider a single-particle state bound 
to a two-dimensional (2-D) square box of depth $\langle V_{\rm GL}({\bf r})\rangle$
and sides of length $\lambda_{\rm CO}$. For Y123 we consider a single-particle 
state bound to a stripe-like 2-D rectangular box of depth 
$\langle V_{\rm GL}({\bf r})\rangle$, width $\lambda_{\rm CO}$, and length 
equivalent to the CDW correlation length, which is much longer than 
$\lambda_{\rm CO}$\cite{Blanco-Canosa2014,Huecker2014,Comin2015a}. We assume in 
both cases that the PG is proportional to the numerical solution of the 
corresponding 2-D Schr\"odinger equation for the ground state binding energy. 
The proportionality factor is estimated using experimental values of the 
pseudogap $\Delta_{\rm PG}$ for Y123 and (Pb, La)-Bi2201 at $p = 0.16$ 
(Tables 1 and 2), and the values of $\Delta_{\rm PG}$ are calculated for the 
underdoped samples using the respective values of $\langle V_{\rm GL}({\bf r})\rangle$  and $\lambda_{\rm CO}$. To further assess the accuracy of the 
results for the underdoped samples, we convert $\Delta_{\rm PG}$ to $T^*$ 
using the experimental ratios $T^*/\Delta_{\rm PG} = 170$ K/76 meV and
$T^*/\Delta_{\rm PG} = 241$ K/ 30 meV for optimally-doped Y123 and (Pb, La)-Bi2201, respectively. As shown in Fig. \ref{fig4} and Table \ref{table2}, the calculated values of $T^*$ 
for (Pb, La)-Bi2201 at p = 0.126 and
0.141 agree well with measurements of the PG onset temperature for La-doped 
Bi2201\cite{Yurgens2003}. Reasonable agreement is also obtained between the calculated and experimental\cite{Daou2010} values of $T^*$ for Y123 at $p = 0.09$ and 0.12 
(Fig. \ref{fig4} and Table \ref{table1}).  
As mentioned in the introduction we presume the onset of CO domains at 
$T^*$ above the long range temperature $T_{\rm CO}(p) <  T^*(p)$, 
a behavior that has been detected by some different experiments\cite{Comin2014,
Comin2015a,DaSilvaNeto2014,Parker2010}.

\begin{table}
\caption{
Experimental data for $\Delta_{\rm SC}$  and $\Delta_{\rm PG}$determined from universal curves that describe Y123 and a number of other high-$T_c$
cuprate superconductors (Ref. \onlinecite{Huefner2007}). An experimental value for 
$\Delta_{\rm SC}$ at $p = 0.12$ is omitted, since
deviations from the universal curve are expected for 
Y123 near 1/8 hole doping, where $T_c$
plateaus. The value of $T^* \approx 278$ K at $p = 0.09$ is estimated from a linear extrapolation of data in
(Ref. \onlinecite{Vig2016}). We tune the scaling factors explained in the text to 
yield the blue and green values at optimum doping. 
Red values are calculated with the same parameters.\label{table1}
}
 \begin{tabular}{|cccc|}\hline \hline
          &         $ p = 0.16 $  & $ p = 0.12 $  &$ p = 0.09 $   \\ \hline
$\lambda_{\rm CO} (a_0)$\cite{Blanco-Canosa2014,Huecker2014,Comin2015a}&$3.49 \pm 0.16$& $3.15 \pm 0.16$ & 
$3.00 \pm 0.16 $\\
$ \langle V_{\rm GL}({\bf r})\rangle$ &  {\color{red} -0.156} &{\color{red}  -0.110 } & {\color{red} -0.105} \\
$\rho_{b}(1.05T_c) (\mu\Omega$ cm)\cite{Segawa2001} & $\approx$ 40  &  $\approx$ 50  
& $\approx$ 70 \\  
\hline
$\Delta_{\rm SC}$\cite{Huefner2007} (meV) & $42\pm 2$  &  -  & $24 \pm 1$ \\
$\langle \Delta_{d}({\bf r})\rangle$ (meV & {\color{blue} 43.8} & {\color{red} 25.5} &
{\color{red} 23.3} \\
$T_c$ (K)\cite{Segawa2001,Daou2010} & 93.4  &   66  &   55 \\
         &{\color{blue} 92.0} & {\color{red} 68.1 }&  {\color{red} 55.6} \\ \hline
$\Delta_{\rm PG}$\cite{Huefner2007} &   76   &   104   & 124 \\
 & {\color{green} 76} & {\color{red} 93.4 $\pm$ 9} & {\color{red} 103.3 $\pm$ 9}\\
 $T^*$\cite{Daou2010}  &   170     &  232   & ~ 278 \\
  &{\color{green} 170} & {\color{red} 209 $\pm$ 19} &{\color{red} 231 $\pm$ 21} \\
\hline \hline
 \end{tabular}
 \end{table}

\begin{table}
\caption{
Experimental data for $\Delta_{\rm SC}$  and $\Delta_{\rm PG}$ 
from STM with the indicated $T_c$ values\cite{Wise2008}. 
The hole doping $p$ was determined from the $T_c$ versus $p$ relationship
obtained by X-ray absorption experiments on (Pb, La)-Bi2201 and La-doped 
Bi2201\cite{Schneider2005}. The in-plane resistivity $\rho_{\rm ab}$ data area
for La-doped Bi2201\cite{Ono2003} and $T^*$ data are from intrinsic tunneling
measurements\cite{Yurgens2003}. We tune the scaling factor explained in the text to 
yield the blue and green values at optimum doping. Red values are 
calculated without any extra parameters. \label{table2}
}
 \begin{tabular}{|cccc|}\hline \hline
          &         $ p = 0.16 $  & $ p = 0.12 $  &$ p = 0.09 $   \\ \hline
$\lambda_{\rm CO} (a_0)$\cite{Wise2008}&$6.2 \pm 0.2$& $5.1 \pm 0.2$ & 
$4.5 \pm 0.2 $\\
$ \langle V_{\rm GL}({\bf r})\rangle$ &  {\color{red} -0.1022} &{\color{red} -0.1018 } & {\color{red} -0.1021} \\
$\rho_{b}(1.05T_c) (\mu\Omega$ cm)\cite{Ono2003} & 18.3  &  24.8 & 28.3  \\  
\hline
$\Delta_{\rm SC}$\cite{Wise2008} (meV) & 15  &  -  &  - \\
$\langle \Delta_{d}({\bf r})\rangle$ (meV & {\color{blue} 15.8} & {\color{red} 13.2} &
{\color{red} 9.2} \\
$T_c$ (K)\cite{Wise2008} & 35  &   32  &   25 \\
         &{\color{blue} 35.2} & {\color{red} 32.5 }&  {\color{red} 24.7} \\ \hline
$\Delta_{\rm PG}$\cite{Wise2009} & 30$\pm$12   & 45$\pm$15   & 68$\pm$20 \\
 & {\color{green} 30} & {\color{red} 44.3} & {\color{red} 54.2}\\
 $T^*$\cite{Yurgens2003}  &   241     &  355   & 446 \\
  &{\color{green} 241} & {\color{red} 327.9 } &{\color{red} 427.4} \\ \hline \hline
 \end{tabular}
 \end{table}
 
\section{Comparison of Calculated and Experimental Parameters}

Tables \ref{table1} and \ref{table2} contain values of experimental parameters (denoted by black text) used in the calculations for 
each compound, and the calculated parameters (denoted by red text). For each cuprate family the calculated values at 
$p = 0.16$ (which are denoted by blue and green text) were multiplied by a scaling factor to match experimental 
values as follows:\\

i) The proportionality constant between $ \langle V_{\rm GL}({\bf r})\rangle$ and the attractive pairing potential 
$V$ of Eq. (\ref{Deltad}) was adjusted to yield a calculated value of $\langle \Delta_{d}({\bf r}) \rangle$ that 
approximately equals the experimental value of the SC gap $\Delta_{\rm SC}$ at $p = 0.16$. 
This proportionality constant, once determined, was subsequently used for all other values of $p$.  \\
ii) The scaling factor between the normal resistance $R_n$ in Eq. (3) and the resistivity $\rho_{\rm b}(1.05Tc)$ 
just above $T_c$ was adjusted until the calculated value of $T_c$ at $p = 0.16$ approximately equaled the experimental value. 
This same scaling factor was used for all other values of $p$.  \\
iii) The ground state binding energy of a single-particle in a 2-D square (rectangular) box was multiplied 
by a proportionality factor so as to equal the PG of (Pb, La)-Bi2201 (Y123) at $p = 0.16$. Again, this same 
proportionality constant was used for the calculations at other dopings.
 
 \section{Fermi arc formation}

Next we show that the phase separation approach considered above is able to 
reproduce the ungapped portion of the Fermi surface (Fermi arcs) that is known 
to occur near the nodal region just above $T_c$\cite{Lee2007,Yoshida2012,Keimer2015}.
We start by recalling that 
Figs. 1(c), 1(f), 2(c) and 2(f) show domains of SC order parameter modulations. 
To each domain we assign a label $j$ and a complex SC order parameter $\Delta_d(k,T)\exp(i\Phi_j)$, where  
$\Delta_d(k,T) = \Delta_0(T)[\cos(k_xa_0) - \cos(k_ya_0)] = \Delta_0(T) 
\cos(2\phi)$, $\phi$ is the azimuthal angle measured from the $x$-direction 
in the CuO$_2$ plane and $\Delta_0(T)$ is the wave function amplitude in the 
j$^{th}$ domain at temperature $T$. The $d$-wave symmetry implies larger 
supercurrents flowing in the CuO$_2$ plane along the antinodal directions 
parallel to the Cu-O bonds, and vanishing values along the nodal directions 
$\phi = \pm \pi/4$ and $\pm \pi/4$. The local intrinsic SC phase $\Phi_j$ and the 
superfluid density $n_j \propto \Delta_d(r_j,T)^2$ are canonically conjugate
variables\cite{Spivak1991}, leading to large fluctuations of the phase $\Phi_j$
along the nodal directions, where $n_j$  and $\Delta n_j$ vanish. This is due to the quantum uncertainty principle and we may write 
$\Delta \Phi_j(\phi) \propto 1/\cos(2\phi)$ to indicate the azimuthal dependence of the phase uncertainty, which has its maximum and minimum values along the nodal and antinodal directions, respectively. Furthermore, $\Delta \Phi_j$ has a clear dependence on the Josephson coupling. In particular, as shown in the previous 
section, at $T < T_c$ all $\Phi_j$ are locked together leading to long range SC order, but at $T > T_c$ phase decoupling occurs because 
$\langle E_{\rm J}(p, T)\rangle < k_{\rm B}T$ and concomitantly 
$\Delta\Phi_j$ increases with $T$ up to the temperature at which 
$\langle E_{\rm J}(p, T)\rangle$ vanishes.  In particular, 
$\Delta \Phi_j$ increases monotonically from near zero at $T_c$ to very large values 
near $T^*$.
Furthermore, $\Delta \Phi_j$ has a large anisotropy when combined with the quantum effects discussed above. The two distinct contributions are separable, such that
$\Delta \Phi_j(p, T, \phi) = \Delta \Phi_j(p, T) \Delta \Phi_j(\phi)$. We drop the index $j$ because $\Delta \Phi_j(\phi)$  is the same for all domains, and assume 
$\langle E_{\rm J}(p, T)\rangle$ is the same for all $j$. These considerations 
imply that just above $T_c$ the electrons ejected by ARPES from different domains come from regions where the SC order parameter has essentially the same $\Phi_j$ 
along the antinodal directions, and $\Delta \Phi_j(\phi) \approx 0 $. 
On the other hand, such phase coherence is lost near the nodal directions 
where $\Delta \Phi_j(\phi) \approx \pi $ is maximum. Consequently,
the average SC amplitude measured by ARPES may be written as follows\cite{DeMello2012},

\begin{flalign}
& \langle \Delta_{d}(p, T, \phi)\rangle = \nonumber \\ 
& \frac{|\cos(2\phi)|}{\Delta \Phi_j(p, T, \phi)}
 \int_{0}^{\Delta \Phi_j(p, T, \phi)} \langle \Delta(p, T)\rangle 
 \cos(\Phi) d\Phi   \nonumber \\ 
&= \langle \Delta(p, T)\rangle |\cos(2\phi)| \frac{1}{\Delta \Phi_j(p, T, \phi)}
 \sin[\Delta \Phi_j(p, T, \phi)] .
 \label{Dcoher}
\end{flalign}

This expression contains the two distinct contributions that weaken phase coherence, 
one from quantum oscillations that depends only on the 
azimuthal angle $\phi$ and one from thermal oscillations that competes with the average Josephson coupling, leading 
to $\Delta \Phi_j(p, T, \phi) = \Delta \Phi_j(p, T) \Delta \Phi_j(\phi)$ for all ``j``. We may take
$\Delta \Phi(\phi) \sim 1/\cos^2(2\phi)$, which satisfies the expected 
inverse cosine dependence and the square makes it symmetric and always positive around the nodal directions 
($\phi = \pm \pi/4$ and $\pm \pi/4$). Now we can infer the functional form of $\Delta \Phi(p, T)$ 
noting that for $T < Tc$, 
all $\Phi_j$ are locked together leading to long range order and $\Delta \Phi_j \sim 0 $. On the other hand, 
for $T \leq T_c$ all $\Phi$ decouple because $\langle E_{\rm J}(p, T)\rangle < k_{\rm B}T$ and 
$\Delta \Phi > 0$. Above $T_c$, $\langle E_{\rm J}(p, T)\rangle $  decreases 
with increasing $T$ and vanishes near $T^*$. Concomitantly $\Delta \Phi$ increases. Thus, there are three distinct 
temperature dependent regimes:\\
(i) $T \leq T_c$: Since $\Delta \Phi \sim 0 $, Eq. (\ref{Dcoher}) is easy to solve and we obtain the ``bare``
expression $\langle \Delta_{d}({\bf r}, T, \phi)\rangle = \langle \Delta_d(p, T)\rangle |\cos(2\phi)$.\\  
(ii) $T > T^*$: $\langle \Delta(p, T)\rangle \sim 0$ and it is clear that there is no gap.\\
(iii) $T_c  < T$: Taking into account the effect of  $\langle E_{\rm J}(p, T)\rangle$ we assume
$\Delta \Phi(p, T) = A[1- \langle E_{\rm J}(p, T)\rangle/k_{\rm B}T_c(p)$  
where $A$ is a constant. This expression vanishes at $T_c$ and increases monotonically with $p$, 
as expected from ARPES experiments\cite{Lee2007,Yoshida2012}. Thus, putting all together,
$\Delta \Phi(p, T, \phi) = [A/\cos^2(2\phi)][1- \langle E_{\rm J}(p, T)\rangle/k_{\rm B}T_c(p)$     and 
we obtain the value of $A$ by comparing with the onset of the measured\cite{Lee2007} gapless region for a given sample. 
To reproduce the measured gapless regions we also assume in Eq. (\ref{Dcoher}) that 
$\langle \Delta_d(p, T, \phi)\rangle \sim 0$
whenever $\Delta \Phi(p, T, \phi) \ge \pi$, 
due to destructive phase interference from electrons ejected from different domains.

\begin{figure}
\includegraphics[height=5.0cm]{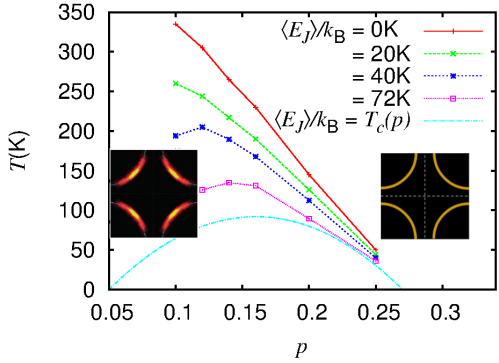}
\caption{
Josephson coupling energy phase diagram and schematic Fermi surface. 
The variation of the average Josephson 
coupling energy with doping and temperature for Bi2212 (from 
Ref. \onlinecite{DeMello2014}).  Above the onset of 
$\langle E_{\rm J}\rangle /k_{\rm B} = 0$ K 
there is no SC gap. The corresponding gapless Fermi surface is depicted at the right of the phase diagram. 
In the region of the phase diagram where $T \ge  T_c$, the average SC gap may be finite. However, combined thermal 
and quantum phase fluctuations may cause destructive interference in the ARPES data along the nodal directions, 
leading to the gapless Fermi arcs shown in the left of the figure.
}
\label{fig5}
\end{figure}

\begin{figure}
\includegraphics[height=5.0cm]{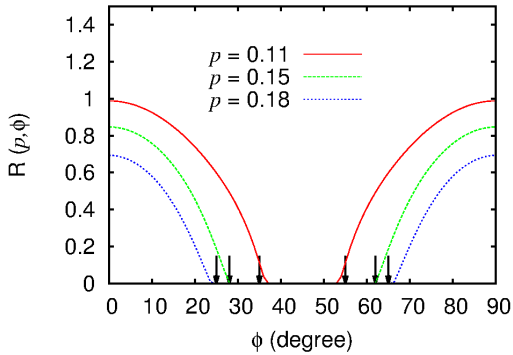}
\caption{The phase factor from the SC fluctuation as function of the azimuthal angle. The envelope of the SC 
amplitude $\langle \Delta_{d}(p, T, \phi)\rangle $ according to Eq. (\ref{Dcoher}). The arrows show the limits 
of the gapless region as determined 
from ARPES experiments on Bi2212 at $T = T_c(p) + 10$ K for three dopings\cite{Lee2007}.
}
\label{fig6}
\end{figure}

Specifically we use the ARPES measurements\cite{Lee2007} at 
$T = 102$ K for the Bi2212 compound with $T_c = 92$ K, 
which shows a gapless region between $28^0 \le \phi \le 62^0$ to derive the constant $A$. 
Equating $\Delta \Phi (p \sim 0.15, T = 102  {\rm K}, \phi = 28^0,  62^0) = \pi$, yields 
$\langle \Delta_{d}(p, T, \phi)\rangle \sim 0$ and this is possible if we
take $A = 2.84\pi$. 
Note that $\phi$ is measured from the $(\pi, \pi)$ to $(0, \pi)$ direction of the Brillouin zone according 
to Refs. \onlinecite{Lee2007} and \onlinecite{Yoshida2012}. With  $\Delta \Phi(p, T, \phi)$ determined, 
we may apply the derived equation to any sample. In particular, we
apply this expression to the other two Bi2212 compounds measured in Ref. \onlinecite{Lee2007}. 
Some above $T_c$ values of $\langle E_{\rm J}(p, T)\rangle $ used in the calculations are plotted in Fig.(\ref{fig5})
for illustration purpose.
Accordingly we obtain for the $T_c = 86$ K 
compound a gapless region at $23.8^0 \le \phi \le 66.2^0$, which compares well with the experimental 
$25^0 \le \phi \le 65^0$. 
For the underdoped $T_c = 75$ K Bi2212 sample at $T = 85$ K, we obtain $36.5^0  \le \phi \le 53.5^0$, which is also 
in good agreement with the experimental result\cite{Lee2007}. 
We summarize the Fermi arc calculations for the three samples in Fig. \ref{fig6}, where the results of the ''envelope''
phase factor of $\langle \Delta_{d}(p, T, \phi)\rangle$ 
\begin{eqnarray}
 R = \frac{|\cos(2\phi)|}{\Delta \Phi(p, T, \phi)}\times \sin[\Delta \Phi_j(p, T, \phi)]
 \label{envelope}
\end{eqnarray}
from Eq. (\ref{Dcoher}) are in good agreement with the experiments\cite{Lee2007}. 
The arrows mark the experimentally determined onset of the
gapless regions for each sample, as described above.

\section{Summary and conclusions}\label{conc}

Our theory infers a fundamental link between the periodicity of the CDW and 
the PG and SC energy scales of high-temperature cuprate superconductors, and 
shows that within this framework it is possible to account for the onset 
temperatures $T_c$ and $T^*$ of two different cuprate families. We stress that 
the only quantitative assumptions made in our calculations for underdoped Y123 and (Pb, La)-Bi2201 pertain to a natural scaling factor, which we have determined by scaling calculated free energy parameters to achieve values of the PG and SC gap 
that agree with experimental values at one particular doping. Our model is general 
in the sense that it can be applied to other cuprate families, provided the doping dependence of the CDW order is known. 

Our approach generates a local free energy potential having a spatial periodicity 
that matches that of the experimentally observed short-range static CDW order.  Our calculations in the framework of BdG theory yield different SC amplitudes in 
distinct charge-ordered domains that generally vanish only above $T_c$. 
Our approach is consistent with experiments\cite{DaSilvaNeto2014,Wise2008,Gomes2007,
Dubroka2011} that measure a finite SC amplitude above $T_c$, and promotes 
the scenario whereby the SC resistive transition marks the onset of global 
phase coherence between SC domains. In our model Fermi arcs appear above $T_c$
because there are large phase fluctuations along the nodal directions where the superfluid density vanishes. The increase of the arcs size with $p$ is 
reproduced because the dependence of $\langle E_{\rm J}(p, T)\rangle$ on 
the temperature changes rapidly with doping.   

Finally we address the experimental observations indicating a competition 
between superconductivity and CO. While X-ray experiments show a decrease of the 
CDW diffraction intensity and correlation length below $T_c$\cite{Chang2012,
Ghiringhelli2012,Blanco-Canosa2014,Huecker2014,Comin2015b}, these measurements 
seem to be detecting static charge correlations. Static CDW order competes 
with superconductivity by reducing the number of charge carriers available for pairing. On the other hand, our theory requires that dynamic CO is also present to induce a fluctuating hole-pair potential that scales with $V_{\rm GL}$. 
While there is some evidence for CDW fluctuations from optical pump-probe\cite{Torchinsky2013} 
and low-energy, momentum-resolved electron energy-loss spectroscopy\cite{Vig2016} 
experiments, there is currently insufficient experimental information 
to assess the pervasiveness or significance of fluctuating CO in cuprates.

\begin{acknowledgments} 
We thank S. Ono and Y. Ando for providing us with their resistivity data. We also thank Andrea
Damascelli, Andr\'e-Marie Tremblay and John Tranquada for informative discussions.
Supported by the Brazilian agencies FAPERJ and CNPq (E.V.L.M.); the Canadian
Institute for Advanced Research (CIFAR) and the Natural Sciences and Engineering
Research Council of Canada (J.E.S.).
\end{acknowledgments}

\appendix 
\section{CHARGE ORDER SIMULATIONS}
  
\begin{figure*}
\includegraphics[height=4.5cm]{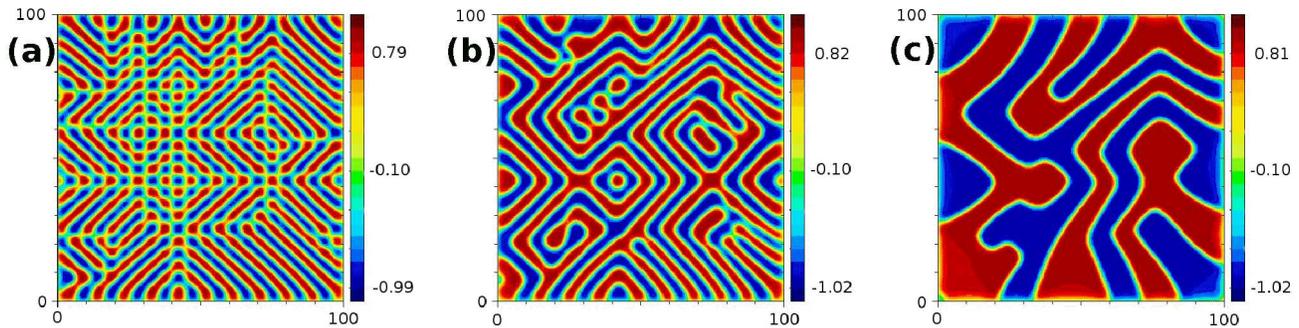}
\caption{ Two-dimensional 
CH simulations of $u(r, t)$ for $\alpha = B = 1$, $\varepsilon = 0.012$ and 
time steps (a) $n = 900$, (b) $n = 1500$, and (c) $n = 9000$. These plots are continuation
of the time evolution of Fig. 2(a) with $n = 700$.
}
\label{figA1}
\end{figure*}

  To describe the growth and development of spatial charge inhomogeneity in the
  CuO$_2$ planes we applied a theory 
of phase-ordering dynamics, whereby the system evolves through domain coarsening when quenched from a homogeneous 
into a broken-symmetry phase. The time-dependent CH approach provides a simple way to determine the time evolution of 
the CO process\cite{Cahn1958}. The CH equation 
can be written in the form of the following continuity equation for the local free energy current 
density $ {\bf J} = -M \nabla^2((\partial f/\partial u)$\cite{Bray1994}

\begin{equation}
 \frac{\partial u}{\partial t} = -{\bf \nabla . J} = -M\nabla^2 \left[ \varepsilon^2\nabla^2 u + 
 \frac{\partial V_{\rm GL}}{\partial u} \right] ,
 \label{CH}
\end{equation}
where $M$ is the charge mobility that sets the phase separation time scale. The order parameter varies between
$u({\bf r},t) \sim 0$ for the homogeneous system above the phase-separation onset temperature $T_{\rm PS}$, and 
$u({\bf r},t \rightarrow \infty) = \pm 1$ for the extreme case of complete phase separation. We
solved the CH equation by a stable and fast finite difference scheme with free boundary
conditions\cite{DeMello2012}. The spatial dependence of the charge density obtained by numerically solving the
CH equation evolves with time $t = n \delta t$, where n is the number of time 
steps $\delta t$. When the order
parameter is conserved, as in phase separation, the charges can only exchange 
locally rather than
over large distances. This leads to diffusive transport of the order parameter. Consequently, at
early times or small n, we obtain charge modulations with periodicities 
of only a few lattice
constants. Using different parameters and initial conditions we 
are able to reproduce the
experimentally determined CO patterns in cuprates. Although these simulations are not the stable
solutions of the CH equation (as is clear from the time evolution of the simulations shown in Fig. \ref{figA1}), 
the aim here is to generate periodic charge modulations with experimentally determined
wavelengths that can be subsequently used to calculate the SC gap and PG in our phase-
separation model. For convergence the time step $ \delta t$ and spatial 
step $h \approx 1/N$ for a square lattice of
$N^2$ sites must be such that $ \delta t \le h^2 /9$ (Ref. \onlinecite{DeMello2012}). For the calculations here we 
used $\delta t \le h^2 /10$ and $h = 1/100$.

In the main paper we present detailed CO, $T_c$ and $T^*$ calculations for six compounds. Three of the Bi2201
and three of the Y123 families. In the next two paragraphs we give the values of
some parameter used in the CH simulations.

{\bf (Pb, La)-Bi2201}: Simulations with $\alpha = B = 1$, time steps of $n = 700, 900, 1300$ and 
$\varepsilon = 0.012, 0.014, 0.0175$ yield checkerboard CO patterns with the desired wavelengths 
$\lambda_{\rm CO}$ for (Pb, La)-Bi2201 (at $p = 0.126, 0.141$ and $0.16$) near $4.5 a_0$, $5.1 a_0$, and $6.2 a_0$, 
respectively. The fewer time steps required to simulate the CO patterns of the underdoped samples is 
indicative of a reduced charge mobility, and is consistent with an increase of the normal-state resistivity. 
At later times (i.e., greater n) the periodic electronic structure evolves into an irregular patch-like 
system of segregated low- and high-charge density regions. In addition, the length scale 
of the system increases with the phase separated regions forming larger domains. 
This latter situation was considered previously\cite{DeMello2014}.  Fig. \ref{figA1} 
shows CH simulations of $u({\bf r}, t)$ 
at times beyond where checkerboard CO with 
$\lambda_{\rm CO}  = 4.5a_0$ is observed in (Pb, La)-Bi2201 at $p = 0.126$.

{\bf Y123}: Simulations with $\alpha$ = 1, $B$ = 5, $\varepsilon$ = 0.0053, 0.0055, 0.0058
and time steps $n = 35, 38, 42$ yield charge stripe patterns with the desired wavevectors 
$Q = 0.333, 0.317$ and 0.287 r.l.u. 
($\lambda_{\rm CO} = 1/Q$) 
estimated from (Refs. \onlinecite{Blanco-Canosa2014,Huecker2014,Comin2015a}) 
for Y123 at $p = 0.09, 0.12$ and 0.16, respectively.  
Note that the values of n are much shorter than needed to simulate the checkerboard CO patterns of (Pb, La)-Bi2201. 
Because of the fewer time steps, the simulations for Y123 are somewhat less sharp.

\section{Combined Bogoliubov-deGennes (BdG) and Cahn-Hilliard (CH) Calculations}

\begin{figure}
\includegraphics[height=5.0cm]{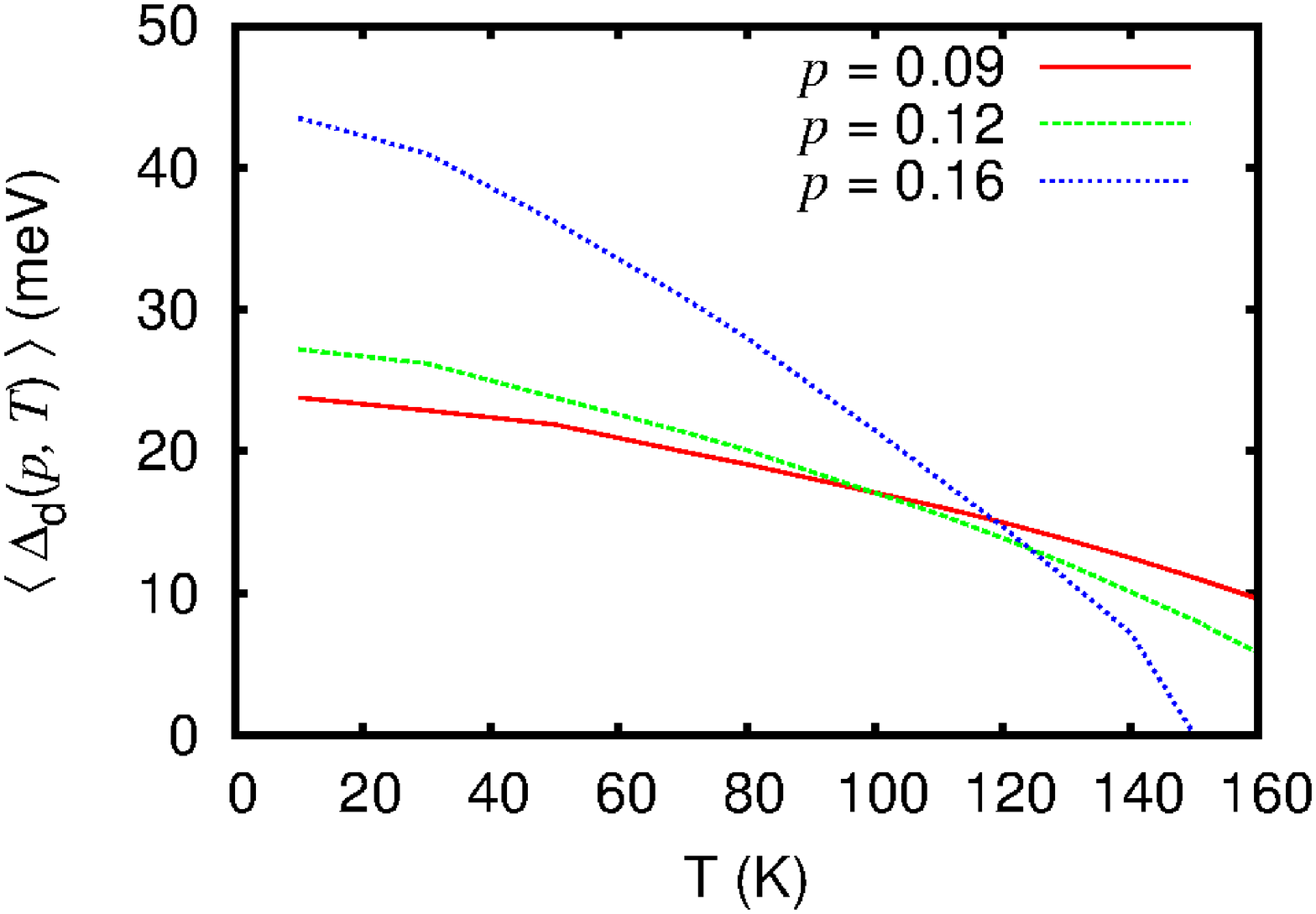}
\caption{Example of calculated $\langle \Delta_{d}(p, T) \rangle$ used to obtain $T_c$: 
The average SC amplitudes for Y123 from BdG Eq. \ref{Deltad} used in the calculations of $\langle E_{\rm J}(p, T)\rangle $ 
[see Eq. \ref{EJ(T)}]. 
The low temperature limits of $\langle \Delta_{d}(p, T = 0) \rangle$ are also listed in Table \ref{table1}. 
}
\label{figA2}
\end{figure}

We performed self-consistent calculations with the BdG theory (Refs.\onlinecite{deMello2004,deMello2009}) 
for each of the CH simulated charge density maps (Figs. \ref{fig1}(a) and 
\ref{fig1}(d), and 
Figs. \ref{fig2}(a) and \ref{fig2}(d)). To calculate the SC pairing
amplitude we assumed the attractive interaction potential $V$ scales with $V_{\rm GL}^{\rm min} \sim \langle 
V_{\rm GL}({\bf r})\rangle$. 
The SC calculations begin with the extended Hubbard Hamiltonian\cite{deMello2004,deMello2009}.
To describe the charge carriers dynamics in the CuO$_2$ planes of the HTSC we consider this Hamiltonian in a square lattice
\begin{eqnarray}
H&=&-\sum_{ \{ ij \} \sigma }t_{ij}c_{i\sigma}^\dag c_{j\sigma}
+\sum_{i\sigma}\mu_i n_{i\sigma}
\nonumber \\
&&
+U\sum_{i}n_{i\uparrow}n_{i\downarrow}-
\frac V 2 \sum_{\langle ij \rangle \sigma
\sigma^{\prime}}n_{i\sigma}n_{j\sigma^{\prime}},
\label{Ham}
\end{eqnarray}
where $c_{i\sigma}^\dag (c_{i\sigma})$ is the usual fermionic creation (annihilation) operator at site $i$, 
the spin $\sigma$ is up $\uparrow$ or down $\downarrow$. $n_{i\sigma} =  c_{i\sigma}^\dag c_{i\sigma}$ 
is the number operator, and $t_{ij}$ is the hopping between sites $i$ and $j$. 
$U$ is the magnitude of the on-site repulsion, and $V$ is the magnitude of the nearest-neighbor attractive 
interaction. $\mu_i$ is the local chemical potential derived in the self-consistent process through which the 
local charge density is calculated by the CH equation and is maintained fixed. 
For (Pb, La)-Bi2201 we used nearest-neighbor hopping $t = 0.15$ eV, next-nearest-neighbor hopping $t_2 = -0.27t$, 
and third nearest-neighbor hopping $t_3 = 0.19t$ derived from angle-resolved photoemission spectroscopy (ARPES) 
dispersion relations\cite{Norman1995}. 
For Y123, we used the ARPES results $t = 0.15$ meV, $t_2 = -0.50t$ and $t_3 = 0.16t$ (Ref. \onlinecite{Schabel1998}). 
The BdG mean-field equations are\cite{deMello2009}

\begin{equation}
\begin{pmatrix} K         &      \Delta_d ({\bf r}) \cr\cr
           \Delta_d^*({\bf r})    &       -K^*
\end{pmatrix}
\begin{pmatrix} u_n({\bf r})      \cr\cr
                v_n({\bf r})
\end{pmatrix}=E_n
\begin{pmatrix} u_n({\bf r})       \cr\cr
                 v_n({\bf r})
\end{pmatrix}
\label{BdGmatrix}
\end{equation}
with 
\begin{eqnarray}
Ku_n({\bf r})&=&-\sum_{\rm {\bf R}}t_{\rm {\bf r,r+R}}u_n({\bf r+R})
+ \mu ({\rm {\bf r}})u_n({\rm {\bf r}})
\nonumber \\
\Delta_d u_n({\bf r})&=&\sum_{\rm {\bf R}}\Delta_d({\rm {\bf r}})u_n({\rm {\bf r+R}}),
\label{g}
\end{eqnarray}
and similar  equations for $v_n({\bf r})$, where {\bf r+R} is the position of the nearest-neighbor
sites, and $\mu ({\rm {\bf r}}) \equiv \mu_i$ is the local chemical potential. 
These equations are solved numerically for eignevalues  $E_n(\ge 0)$ self-consistently with the spatially-varying 
$d$-wave pairing potential\cite{deMello2004} 
\begin{eqnarray}
\Delta_{d}({\bf r})=&-{\frac V 2}\sum_n[u_n({\bf r})v_n^*({\rm {\bf r + R}})
\nonumber \\
&
+v_n^*({\bf r})u_n({\rm {\bf r+R}})]\tanh{\frac {E_n} {2k_BT}},
\label{Deltad}
\end{eqnarray}
where $V = V(T,p)$ was defined in Eq. \ref{V(T)}. The results of $\langle \Delta_{d}({\bf r}, T) \rangle$ are plotted in
Fig. \ref{figA2} for the three compounds of the Y123 system. Concomitantly, the spatially-varying
hole density of charge carriers is given by
\begin{eqnarray}
p({\bf r}) = 1-2\sum_n[|u_n({\bf r})|^2f_n+|v_n({\bf r})|^2(1-f_n)], 
\label{dop}
\end{eqnarray}
where $f_n = [\exp(E_n/k_{\rm B}T + 1]^{-1}$ is the Fermi occupation function. 
It is important to emphasize that the spatially inhomogeneous distribution of charge generated by the CH equation for 
different dopings was kept fixed while the local chemical potential $\mu ({\bf r})$ was self-consistently determined 
in the convergence process.  This procedure incorporates the charge inhomogeneity in the calculations in a natural way.

%

\end{document}